\documentclass[preprint,12pt]{elsarticle}
\usepackage{graphicx} 
\usepackage[usenames]{color} 
\usepackage{ulem}
\usepackage{amssymb}
\usepackage{multirow}

\journal{Physica A}

\begin{document}

\begin{frontmatter}

\title{Thin film growth models with long surface diffusion lengths}

\author{Tung B. T. To}
\ead{tungto@if.uff.br}
\author{Vitor B. {de Sousa}}
\ead{vitorbds@hotmail.com}
\author{F\'abio D. A. Aar\~ao Reis}
\ead{reis@if.uff.br}
\address{Instituto de F\'{i}sica, Universidade Federal Fluminense,
Avenida Litor\^{a}nea s/n, 24210-340 Niter\'{o}i, RJ, Brazil}

\begin{abstract}
In limited mobility (LM) models of thin film deposition, the final position of each
atom or molecule is chosen according to a set of stochastic rules before the incidence of
another atom or molecule.
Here we investigate the possibility of a LM model to reproduce features of a more realistic approach
that represents the interplay of collective adatom diffusion and the external flux.
In the LM model introduced here, each adatom may execute $G$ hops to neighboring columns of the deposit,
but a hop attempt from a site with $n$ lateral neighbors has probability $P^n$, with $P<1$.
These rules resemble those of the Clarke-Vvedensky (CV) model without energy barriers at step edges,
whose main parameters are the diffusion-to-deposition ratio $R$ on terraces
and the detachment probability $\epsilon$ per lateral neighbor.
At short times, the roughness of the LM model can be written in terms of a scaling function of
$G$ and $P$ and the growth exponent is consistent with the Villain-Lai-Das Sarma universality class.
The evolution of the surface roughness and of the autocorrelation function of the
CV model is reproduced with reasonable accuracy by the LM model with suitable choices
of parameters.
The increase of the parameters $G$ and $R$ of those models produces smoother film surfaces,
while the increase of $P$ and $\epsilon$ smoothen the terrace boundaries at short lengthscales.
However, the detachment probabilities of the two models have very different effects on the
surface roughness: in the LM model, for fixed $G$, the surface roughness increases as $P$ increases;
in the CV model, the surface smoothens as $\epsilon$ increases, for fixed $R$.
This result is related to the non-Markovian nature of the LM model, since the diffusivity of an adatom
depends on its history at the film surface and may be severely reduced after a detachment from a terrace step;
instead, in a collective diffusion model, the detached adatom has the same mobility
as a freshly deposited adatom in the same environment.
This interpretation is supported by the correlation between the surface roughness
and the average number of hops after the last detachment from lateral neighbors in the LM model.
We conclude that, although a limited mobility model can reproduce morphological properties
of a collective diffusion model, the role of apparently equivalent parameters may be
very different, which have consequences for their physical interpretation.

\end{abstract}

\begin{keyword}
thin films \sep growth models \sep surface diffusion \sep dynamic scaling

\end{keyword}

\end{frontmatter}

\section{Introduction}
\label{intro}

Thin film growth models have been intensively studied in the last three decades
due to a large number of applications and due to their fundamental interest
in non-equilibrium statistical mechanics \citep{barabasi,krug,pimpinelli,michely}.
The most realistic models describe thermally activated microscopic processes
such as diffusion, aggregation, desorption, and chemical reactions
which simultaneously occur on the surface of the growing film;
energy barriers for diffusion across step edges and various ranges of interactions
between adatoms may be considered.
The comparison between the parameters used in simulations of these models and
those predicted by ab initio methods (e. g. density functional theory) may
improve the description of a growth process and help the design of films with
the desired properties \citep{etb}.
Among the lattice models of thin film growth \citep{etb}, one of the simplest examples is the
Clarke-Vvedensky (CV) model \cite{cv}, in which the diffusion coefficients have Arrhenius
forms and the energy barriers represent interactions with the neareast neighbors in the lattice.
Extensions of the original CV model have applications to molecular beam epitaxy
of metals or semiconductors \citep{barabasi,krug,pimpinelli,michely,etb,einax},
vapor deposition of organic molecules, and deposition of colloidal particles
\citep{ganapathy,clancy2011,bommel2014,kleppmann2017}.

Thin film growth may also be described with reasonable approximation
by limited mobility models, in which the final aggregation position
of each deposited atom or molecule is chosen according to a set of stochastic
rules before the adsorption of another atom or molecule.
Some well-known examples are ballistic deposition \citep{vold}, the
restricted solid-on-solid model \citep{kk}, and the models for
molecular beam epitaxy of Das Sarma and Tamborenea (DT) \citep{dt}
and of Wolf and Villain (WV) \citep{wv}.
Due to their simple growth rules, they are suitable for kinetic
roughening studies, which require simulation of large systems and long times.
Those studies provide estimates of quantities such as scaling
exponents, height, and roughness distributions, which allow the connection
between the lattice models and hydrodynamic growth equations that
define universality classes \citep{barabasi,krug}.
However, an important question is whether limited mobility models with rules that mimic surface
diffusion can be related to more realistic approaches such as the CV model.

A limited mobility model with this feature was studied in Ref. \protect\cite{CDLM} and was
subsequently termed lateral aggregation of diffusing particles (LADP) \citep{vlds2013}.
In that model, the incident atom can execute a maximal number of $G$ of hops on terraces and
permanently aggregates if it has one or more lateral neighbors.
For a suitable choice of parameters, its roughness scaling is the same as that
of the CV model with irreversible attachment of adatoms to lateral neighbors;
this is a case in which the CV dynamics do not obey detailed balance conditions.
The DT, WV, and large curvature models with long diffusion lengths were also studied in several
previous works \citep{dassarmaPRE2002,chatraphorn2002,disrattakit2016,disrattakit2017},
but no relation with models of collective adatom diffusion was proposed.
An extension of the LADP model in which atoms with more than one neighbor can move was
recently used to describe electrodeposition, in which porous films are formed \citep{edcross}.

In this work, we introduce a limited mobility model which is designed to represent mechanisms
of surface diffusion similar to the CV model:
each adatom can execute a maximum of $G$ hops, detachment from 
lateral neighbors is allowed with probabilities depending on its current neighborhood,
and solid-on-solid conditions are considered (i. e. the films are not porous).
For simplicity, it is hereafter termed LM model.
We determine the effects of its parameters on the scaling of the surface roughness and address
the question of the equivalence with the original CV model.
LM and CV models belong to the universality class of the nonlinear molecular beam epitaxy
equation of Villain, Lai, and Das Sarma (VLDS) \citep{villain,laidassarma}
and we show that it is possible to matching the surface roughness and the autocorrelation
function of the two models by a suitable choice of their parameters.
However, significant differences are observed in the effects of the probabilities of adatoms
detaching from lateral neighbors, which affect the physical interpretation of the LM model.
These differences are discussed, with particular emphasis on the effective time of surface diffusion.

The rest of this paper is organized as follows. In Sec. \ref{basic}, we present the models,
the related growth equation, and the main quantitites to be measured.
In Sec. \ref{roughness}, we analyze the surface roughness in the LM model.
In Sec. \ref{comparison}, we compare results of the LM model with those of the CV model.
In Sec. \ref{detachment}, the effects of the detachment from lateral neighbors in the LM model are
analyzed.
In Sec. \ref{conclusion}, we present our conclusions.

\section{Basic concepts and definitions}
\label{basic}

\subsection{Models}
\label{models}

All models studied here are defined in a simple cubic lattice, with an initially flat
substrate at $z=0$.
The edge of a site is taken as the length unit.
The lateral size of the lattice is denoted as $L$ and periodic boundary conditions are
considered in the $x$ and $y$ directions.
The set of adatoms with the same $\left( x,y\right)$ position is a column of the deposit;
the height variable $h\left( x,y\right)$ is the maximal height of an adatom in that column
(or the number of atoms in the column).

\subsubsection{CV model}
\label{CVmodel}

\begin{figure}[!h]
\begin{center}
\includegraphics [width=0.6\textwidth]{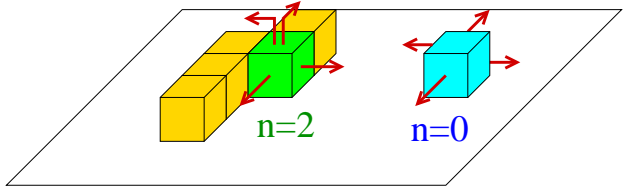}
\caption{Possible hop directions (red arrows) of adatoms with $n=0$ (blue), $n=1$ (brown),
and $n=2$ (green) lateral neighbors; other adatoms are indicated in yellow color.
}
\label{hops}
\end{center}
\end{figure}

Deposition occurs with a flux of $F$ atoms per column per unit time, in the $z$
direction towards the substrate.
An incident atom is adsorbed at the top of a randomly chosen column.
All adatoms at the top of the $L^2$ columns are mobile.
The hopping rate of an adatom with no lateral neighbor is
\begin{equation}
D_0=\nu\exp{\left( -E_s/k_BT\right)}
\label{defD0}
\end{equation}
where $\nu$ is a frequency, $E_s$ is an activation energy, and $T$ is the temperature.
If an adatom has $n$ lateral neighbors, its hopping rate is
\begin{equation}
D=D_0\epsilon^n \qquad , \qquad \epsilon \equiv \exp{\left( -E_b/k_BT\right)} ,
\label{defD}
\end{equation}
where $E_b$ is a bond energy.
Thus, $\epsilon$ may be interpreted as a detachment probability per lateral neighbor.
The hop direction is randomly chosen among the four nearest neighbor (NN) columns
($\pm x$, $\pm y$) and the adatom moves to the top of that column.
Fig. \ref{hops} illustrates the possible hops of three adatoms.

In the original CV model, $\nu=2k_BT/h$, where $h$ is the Planck's constant,
as predicted by transition state theory \cite{cv}.
However, it is more frequent that a constant value $\nu\sim {10}^{12}s^{-1}$ is used
in simulation and analytical works \citep{etb}, and this is also assumed here.
The values of $E_s$ and $E_b$ are determined by material properties, while
$T$ and $F$ are controlled by the deposition technique.

A diffusion-to-deposition ratio is defined as
\begin{equation}
R \equiv \frac{D_0}{F} = \frac{\nu}{F} \exp{\left( -E_s/k_BT\right)} .
\label{defR}
\end{equation}
This is usually interpreted as the number of hops of an adatom in a terrace before
it is buried by the next atomic layer.
In this work, we consider $R$ and $\epsilon$ as the independent parameters of the CV model.

Our simulations are performed in substrates with lateral size $L=1024$.
The simulation time $t$ is expressed as the number of deposited monolayers, which corresponds
to $F=1$.
The deposition is limited to $6\times {10}^3$ monolayers, which is a typical
value for thin films with thicknesses $1$-$3\mu$m.
The lattice size is sufficiently large to avoid finite-size effects.
We consider the ranges of parameters $10\leq R\leq {10}^4$ and $0\leq \epsilon\leq 0.1$;
this is justified because larger values of $R$ produce films with almost flat surfaces
and large values of $\epsilon$ represent solids close to the melting points.
Average quantities are obtained from $100$ independent realizations for each set of parameters.

\subsubsection{LM model}
\label{LMmodel}

In this model, the surface diffusion of each adatom is executed before the next
atom is deposited.
The adsorption of an incident atom occurs at the top of a randomly chosen column.
The diffusion consists of a sequence of $G$ hops to the top of randomly chosen NN columns.
Each hop is executed with probability
\begin{equation}
P_{hop}=P^n ,
\label{Phop}
\end{equation}
where $n$ is the number of lateral neighbors; with probability $1-P_{hop}$, the hop attempt
is rejected.
Thus, $P$ is a probability of detachment per lateral neighbor. 

The illustration of hop directions in Fig. \ref{hops} is also applicable here,
although only one adatom is mobile at each time in the LM model.
Moreover, the detachment probability proposed in Eq. (\ref{Phop}) has
the same form as that of the CV model [Eq. (\ref{defD})].
The LM model is consequently designed with long diffusion lengths and
detachment rates that resemble the dynamics of the CV model.

A closer connection might be possible if we assumed that $P$ has an activated form
as $P \equiv \exp{\left( -E_P/k_BT\right)}$, where $E_P$ is a bond energy.
However, in Sec. \ref{roughness} we will show that $P$ and $\epsilon$ have different
effects on the surface roughness, which prevents the interpretation of $P$ as a
temperature-like parameter.

The LADP model is the case $P=0$ \citep{CDLM}.
It was already used in studies of apparent anomalies of VLDS models \citep{vlds2013} and
to simulate effects of surface diffusion in grain coarsening \citep{reisPRE2017}.
The electrodeposition model of Ref. \protect\cite{edcross} is also similar to the LM,
but it considers a diffusive flux of incident atoms and the adatom hops allow
formation of surface overhangs and, consequently, of porous deposits.
The LM model defined here ($P>0$) was not studied in previous works.

The simulations of the LM model are performed in substrates with lateral size $L=1024$ and
the ranges of parameters $10\leq G\leq 200$ and $0\leq P\leq 0.1$.
$6\times {10}^3$ monolayers are deposited for each set of parameters and
average quantities are obtained from $100$ realizations.
The time unit is also that of deposition of one monolayer.

\subsection{Roughness, correlations, and dynamic scaling}
\label{dynamicscaling}

The surface roughness is defined as
\begin{equation}
W(L,t)\equiv {\left[ { \left<  \overline{{\left( h - \overline{h}\right) }^2}  \right> }
\right] }^{1/2} ,
\label{defw}
\end{equation}
where the overbars indicate spatial averages
and the angular brackets indicate configurational averages.
In a kinetic roughening process at relatively short times, the effects of the finite size of the
substrate are negligible and the roughness increases as 
\begin{equation}
W \sim t^{\beta} ,
\label{defbeta}
\end{equation}
where $\beta$ is called the growth exponent.
This is termed the growth regime.
In the times simulated in this work, the CV and the LM model are in their growth regimes.

The autocorrelation function is defined as \citep{zhao}
\begin{equation}
\Gamma\left( s,t\right) \equiv  \frac{ \left\langle{\left[ \tilde{h}\left( {\vec{r}}_0+\vec{s},t\right)
\tilde{h}\left( {\vec{r}}_0,t\right) \right]}^2\right\rangle }{W^2} \qquad ,\qquad s\equiv |\vec{s}| ,
\label{defcorr}
\end{equation}
where $\tilde{h}\equiv h-\overline{h}$.
The configurational averages are taken over different initial positions ${\vec{r}}_0$,
different orientations of $\vec{s}$ (substrate directions), and different deposits.
This definition implies $\Gamma\left( 0,t\right)=1$ and it is generally expected that
$\Gamma<1$ for $s>0$.

The correlation length $\xi$ may be defined from the correlation function features in
different ways.
In Ref. \protect\cite{zhao}, the length $\xi_e$ is the first value of $s$ in which
$\Gamma\left( s,t\right)=e^{-1}\approx 0.3679$.
A correlation length $\xi_0$ may be defined as the first zero of the
correlation function, $\Gamma\left( \xi_0,t\right)=0$; this may be particularly useful
to characterize surfaces with mounded structure.
In all cases, $\xi$ is a typical length with significant height fluctuations.
In the growth regime, it is expected that $\xi \sim t^z$, where $z$ is the dynamical exponent.

When growth is dominated by surface diffusion, it is expected to be
described by a fourth order stochastic equation in the hydrodynamic limit \citep{barabasi}:
\begin{equation}
{{\partial h(\vec{r},t)}\over{\partial t}} = \nu_4{\nabla}^4 h +
\lambda_{4} {\nabla}^2 {\left( \nabla h\right) }^2 + \eta (\vec{r},t) ,
\label{vlds}
\end{equation}
where $h(\vec{r},t)$ is the height at position $\vec{r}$ and time $t$ in a
$d$-dimensional substrate, $\nu_4$ and $\lambda_{4}$ are constants and $\eta$
is a Gaussian, nonconservative noise [the contribution of the average
external flux is omitted in Eq. (\ref{vlds})].
The linear version ($\lambda_{4}=0$) is the Mullins-Herring (MH) equation \citep{mh} and
the nonlinear version ($\lambda_{4}\neq 0$) is the VLDS equation \citep{villain,laidassarma}.
In $2+1$ dimensions, the best numerical estimates of scaling exponents of the VLDS class
are $\beta\approx 0.20$ and $z\approx 3.3$ \citep{crsosreis,odor2010,xiaSS2013,carrascoVLDS2016}.
They agree with the two-loop renormalization estimates $\beta\approx 0.199$ and $z\approx 3.36$ 
\citep{janssen1997}.

The CV model is described by the VLDS equation in the continuous limit, as shown by
renormalization studies \citep{hasel2007,haselPRE2008} and by numerical calculation of
scaling exponents (growth, dynamical, and the roughness exponent $\alpha=\beta z$)
\citep{kotrla1996,lealJPCM,cv2015,mozo2017}.
For the values $\epsilon\leq 0.1$ studied here and with no energy barrier at terrace steps,
no crossover or instability is expected in the CV model at long times;
for $\epsilon\sim 0.25$ or larger, the system without deposition is close to a roughening transition,
and instabilities appear when step barriers are included, but these phenomena will not be addressed here.
The exponents of the LADP model ($P=0$) were also shown to be consistent with the VLDS
class \citep{CDLM}.

\section{Surface roughness in the LM model}
\label{roughness}

\begin{figure}[!h]
\begin{center}
\includegraphics [width=\textwidth]{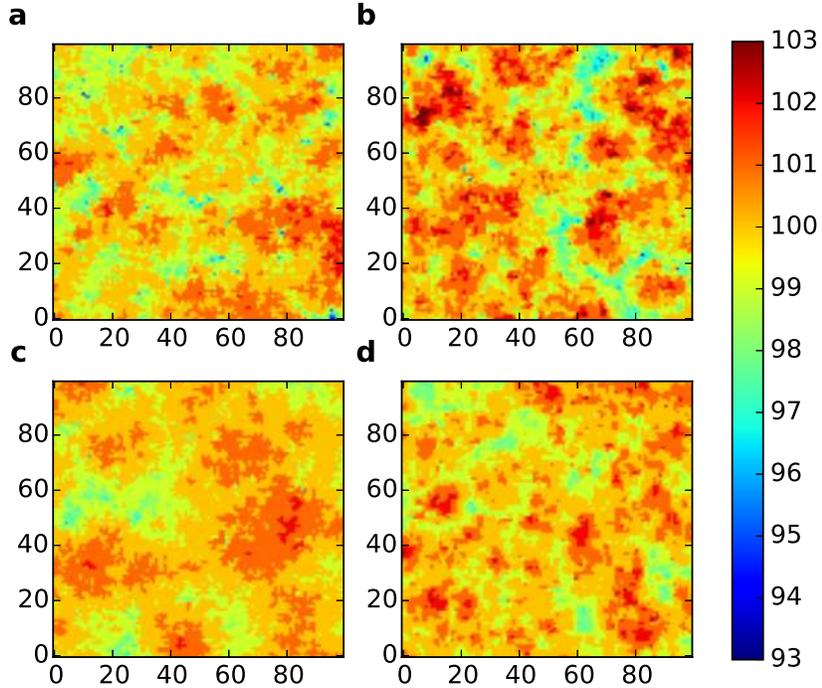}
\caption{Top views of films deposited with the LM model at $t=100$ with parameters
(a) $G=20$, $P=0$; (b) $G=20$, $P=0.1$;
(c) $G=60$, $P=0$; (d) $G=60$, $P=0.1$.
}
\label{topview}
\end{center}
\end{figure}

Figs. \ref{topview}(a)-(d) show top views of films deposited with two values of $G$
and two values of $P$.
These images have lateral size $100$, which is much smaller than the total size of the deposits,
but is suitable for visualization of small lengthscale features.

For $P=0$, the terraces have highly disordered boundaries which resemble
diffusion-limited aggregates \citep{witten}; this morphology is observed since the
early stages of deposition, when a submonolayer is being formed \citep{etb}.
The average terrace size clearly increases with $G$.

For $P>0$, careful inspection of the images in Figs. \ref{topview}(b) and \ref{topview}(d)
show terrace boundaries with some straight regions, particularly for $G=60$.
Such effect is expected because the detachment of weakly bonded adatoms at terrace borders
permits their diffusion until reaching more stable positions, i. e. positions with larger 
numbers of lateral neighbors.
However, the straight regions seldom have more than $10$ adatoms, so the terrace boundaries
are highly disordered in longer lengthscales.

\begin{figure}
\begin{center}
\includegraphics [width=0.7\textwidth]{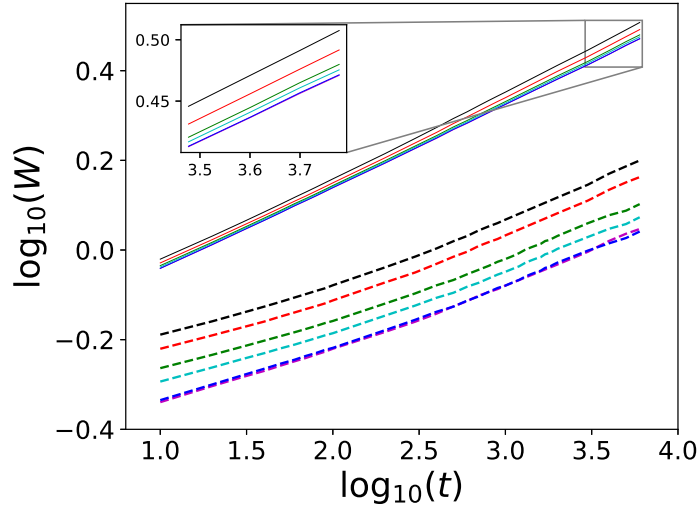}
\caption{Roughness as a function of time of the LM model with $G=10$ (solid lines)
and $G=100$ (dashed lines).
From bottom to top, the values of $P$ are $0$, $0.001$, $0.01$, $0.02$, $0.05$,
and $0.1$.
}
\label{rugLM}
\end{center}
\end{figure}

Figs. \ref{rugLM} show the evolution of the surface roughness of the LM model
with two values of $G$ and several values of $P$.
For fixed $P$, the increase in $G$ has a smoothening effect;
for $G=100$, $W\lesssim 1$ at all simulated times, which means that the film surfaces
have typical fluctuations smaller than the size of an adatom.
This explains why large values of $G$ were not considered here.
For fixed $G$, we observe that the increase in $P$ leads to
an increase in the roughness, even for very small values of this parameter.
This is a surprising effect because increasing $P$ has a smoothening effect on
the terrace borders.

Note that there is no inconsistency in the effects of $P$ on the roughness and on the terrace
boundaries.
The main contribution to the roughness is given by the long wavelength fluctuations in the
vertical ($z$) direction, while fluctuations in terrace borders are measured in small lengthscales
in the directions parallel to the substrate ($x$ and $y$).

The dynamic scaling relation (Family-Vicsek relation \citep{fv}) can incorporate the non-universal
parameters $G$ and $P$.
The dependence of the roughness on $G$ is predicted by a scaling approach \citep{CDLM,cv2015} and
the incorporation in the scaling relation follows the same lines of works on competitive
growth models \citep{horowitz2006,chou2009}.
However, there is no such approach to predict the dependence on $\epsilon$, thus our procedure is to
search for the simplest possible relation that fits the numerical data; this procedure was adopted
in the study of the CV model in Ref \protect\cite{cv2015}.

First, we expect that the roughness scales as $W\sim t^{0.2}/G^{0.5}$ for
$P=0$, as shown by scaling arguments in Ref. \protect\cite{CDLM}.
This means that the scaled time variable need to have the form $t/G^{5/2}$.
Second, the plots of the roughness for fixed $G$ in Fig. \ref{rugLM} suggest an
approximately linear increase of the roughness with $P$, for fixed time.
This leads to the proposal
\begin{equation}
W \approx f\left[ \frac{t}{G^{5/2}} {\left( G^\delta P+b \right)}^\gamma \right] ,
\label{scalingW}
\end{equation}
where $f$ is a scaling function and $\delta$, $\gamma$, and $b$ are constants.
The apparent linearity of the roughness with $P$ suggests that $\gamma\approx 1$, but we do not have
any theoretical explanation to this result.
We also do not have any argument to anticipate the values of $\delta$ and $b$.

\begin{figure}
\begin{center}
\includegraphics [width=0.7\textwidth]{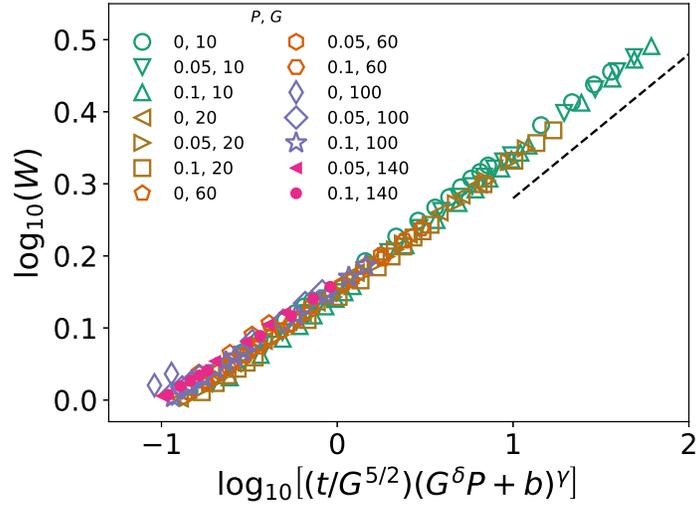}
\caption{Roughness as a function of the time scaled according to Eq. (\ref{scalingW}),
with $\delta = 1.3$, $\gamma\approx 0.9$, and $b=2.5$.
The dashed line has slope $0.20$.
}
\label{collapse}
\end{center}
\end{figure}

Fig. \ref{collapse} shows $W$ as a function of the scaling variable of Eq. (\ref{scalingW}),
for $\delta = 1.3$, $\gamma=0.9$, and $b=2.5$; these are the values
which provide the best collapse of the data into a single curve.
The slope at the longest times is very close to $0.20$, which agrees with the VLDS class.
The proposed scaling form also confirms the general trend of $W$ to increase with $G$
(for fixed $P$) and to increase with $P$ (for fixed $G$).

\section{Comparison between LM and CV models}
\label{comparison}

\begin{figure}
\begin{center}
\includegraphics [width=0.8\textwidth]{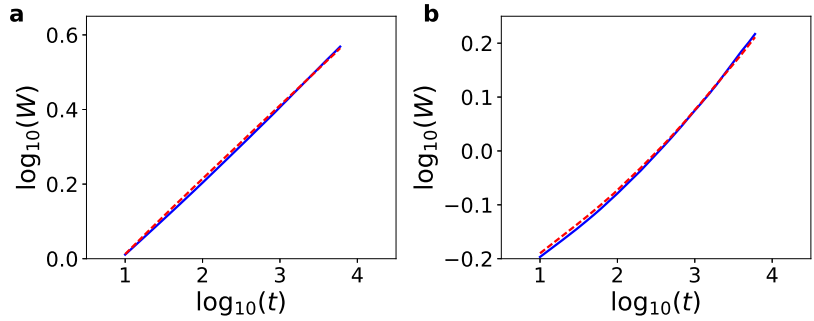}
\caption{Comparison of the roughness evolution of the LM and CV models, with:
(a) $R={10}^2$, $\epsilon=0.05$ (solid line), $G=7$, $P=0.05$ (dashed line);
(b) $R={10}^3$, $\epsilon=0.1$ (solid line), $G=89$, $P=0.1$ (dashed line).
}
\label{Wcomp}
\end{center}
\end{figure}

Since the LM and the CV models have VLDS scaling [Eq. (\ref{defbeta})] with small corrections
at short times, we expect that for some sets of parameters they show
the same evolution of the surface roughness.
For $\epsilon=P=0$, this occurs if $G\approx 0.29 R^{0.6}$ \citep{CDLM}.
Figs. \ref{Wcomp}(a)-(b) show the matching of the roughness evolution for $\epsilon=P>0$
with a suitable choice of $G$ for each values of $R$.
This matching is also possible for $\epsilon\neq P$ with suitable choices of $R$ and $G$.

\begin{figure}
\begin{center}
\includegraphics [width=\textwidth]{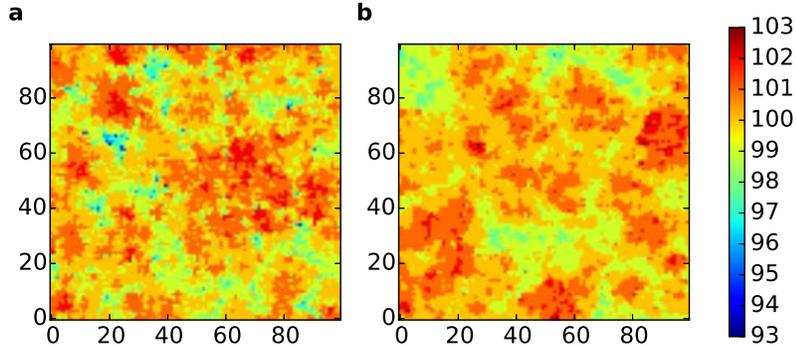}
\caption{Top views of the surfaces grown with the CV model with:
(a) $R={10}^2$, $\epsilon=0.05$ and (b) $R={10}^3$, $\epsilon=0.1$.
}
\label{topviewCV}
\end{center}
\end{figure}

Figs. \ref{topviewCV}(a)-(b) show top views of the film surfaces grown with the CV model
with the same parameters of Figs. \ref{Wcomp}(a)-(b).
The average terrace size increases with $R$, which parallels the increase with $G$ in the LM
model [Figs. \ref{topview}(a)-(d)].
Terrace boundaries are also very rough.
For constant $R$, the increase of $\epsilon$ leads to the formation of a larger number of straight
regions in the terrace borders.
Indeed, the detachment of an adatom from a position with
small number of lateral neighbors permits that it moves until reaching a point with higher
coordination, where it remains for a long time and may be eventually covered by a new atomic layer.
Thus, the effects of $P$ and $\epsilon$ on terrace boundaries are also similar.
However, increasing the detachment probability $\epsilon$ has a smoothening effect \citep{cv2015},
which contrasts with the roughening effect of $P$.

\begin{figure}[!h]
\begin{center}
\includegraphics [width=0.9\textwidth]{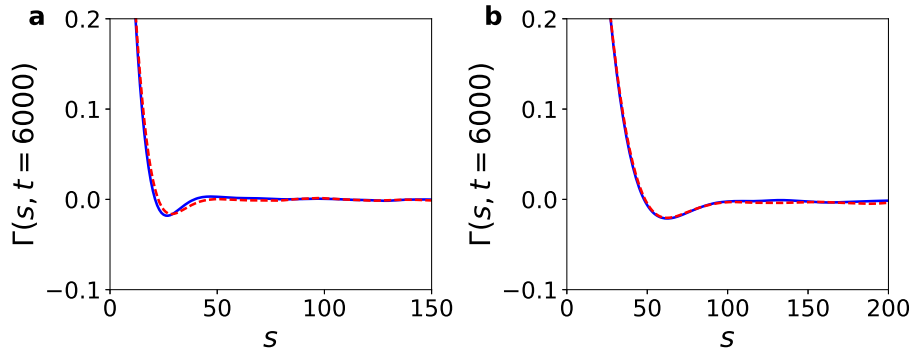}
\caption{Comparison of the correlation function of the LM and CV models at $t=6000$
for the same parameters as in Fig. \ref{Wcomp}(a)-(b).
}
\label{Gammacomp}
\end{center}
\end{figure}

The comparison of the correlation functions of the surfaces with the same roughness
grown by the two models confirm that those surfaces are similar.
Figs. \ref{Gammacomp}(a)-(b) show $\Gamma\left( s,t=6000\right)$ as a function of the distance
$s$, for the same model parameters of Figs. \ref{Wcomp}(a)-(b).
For small $s$, the curves of the two models usually collapse, which means that the estimates
of the correlation length $\xi$ are approximately the same.
There may be difference of a few percents near the first minimum, as shown in Fig. \ref{Wcomp}(a),
which leads to small differences in the correlation length $\xi_0$.
There are other parameter sets [$\left( G,P\right)$ and $\left( R,\epsilon\right)$] which produce
the same roughness evolution and autocorrelation functions with small differences.

In Ref. \protect\cite{cv2015}, a scaling relation analogous to Eq. (\ref{scalingW}) was obtained
for the CV model in the growth regime:
\begin{equation}
W \sim {\left[ \frac{t}{R^{3/2}\left( \epsilon + a\right)} \right]}^{0.2} ,
\label{WCV}
\end{equation}
where $a=0.025$ is a fit parameter.
This relation is consistent with VLDS scaling.
The combination of this result and Eq. (\ref{scalingW}) imply that the same roughness scaling
at short times is obtained in the CV and LM models under the condition
\begin{equation}
R \sim \frac{G^{5/3}}{ {\left( \epsilon + a\right)}^{2/3} {\left( G^\delta P+b \right)}^{2\gamma / 3} } .
\label{Req}
\end{equation}
This relation is presented here for completeness, but it does not add any information on the physics of
the models because it involves four fit parameters ($a$, $b$, $\delta$, and $\gamma$).
However, since the effect of $\epsilon$ is small, it may be helpful to predict the order of magnitude of
the parameter $R$ whose results can be approximated by this LM model.

\section{The effects of detachment probabilities in the LM model}
\label{detachment}

\subsection{Numerical results}
\label{averagenumbers}

The nontrivial effect of the detachment probability $P$ on the roughness is also
illustrated by the comparison of surfaces with the same roughness grown by the LM model
with different parameters.
In Table \ref{table1}, we show the parameters $G$ and $P$ which give three chosen
values of $W$ at $t=6000$.
As $P$ increases from $0$ to $0.1$, the values of $G$ which provide the same
roughness may increase by a factor close to $3$.

\begin{table}[!h]
\centering
\caption{The values of the parameters $G$ and $P$ which provide the given values of
the roughness at time $t=6000$ and the corresponding average number of hops of adatoms since the
adsorption and since the last detachment.}
\label{table1}
\begin{tabular}{|l|l|l|l|l|}
\hline
$W\left( t=6000\right)$ & $G$  & $P$    & $\langle N\rangle$ & $\langle N_{last}\rangle$ \\ \hline
\multirow{3}{*}{$2.30$}  & $16$ & $0$ & $1.979(1)$ & $1.979(1)$ \\ \cline{2-5} 
                        & $20$ & $0.05$ & $2.6925(5)$ & $1.8865(5)$ \\ \cline{2-5} 
                        & $24$ & $0.1$ & $3.672(1)$ & $1.7225(5)$ \\ \hline
\multirow{3}{*}{$1.73$}  & $29$ & $0$    & $3.333(1)$ & $3.333(1)$ \\ \cline{2-5} 
                        & $48$ & $0.05$ & $5.866(1)$ & $3.147(1)$ \\ \cline{2-5} 
                        & $68$ & $0.1$  & $9.595(2)$ & $2.659(1)$ \\ \hline
\multirow{3}{*}{$1.63$}  & $33$ & $0$ & $3.727(2)$ & $3.727(2)$ \\ \cline{2-5} 
                        & $60$ & $0.05$ & $7.138(2)$ & $3.497(1)$ \\ \cline{2-5} 
                        & $89$ & $0.1$ & $12.228(2)$ & $2.874(1)$ \\ \hline
\end{tabular}
\end{table}

We measured the average number of hops executed by an adatom, $\langle N\rangle$,
for each parameter set.
This may be interpreted as the average diffusion time of the adatom since its adsorption,
but excluding the time attached to lateral neighbors.
The values of $\langle N\rangle$ are shown in Table \ref{table1}.
They clearly have remarkable differences for the films with the same roughness.

\begin{figure}[!h]
\begin{center}
\includegraphics [width=0.6\textwidth]{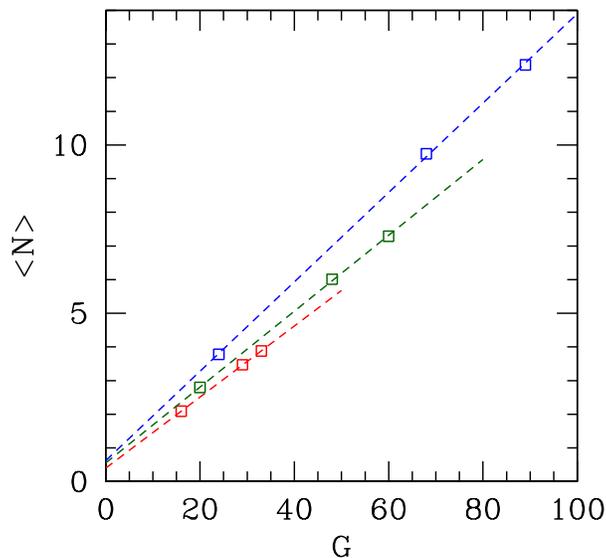}
\caption{Average numbers of hops (total and after the last detachment)
as a function of $G$ for three values of $P$.
The dashed lines are least squares fits of $\langle N\rangle$ for each $P$, with slopes $0.103$, $0.112$, and
$0.132$ from bottom to top.
}
\label{Nmedio}
\end{center}
\end{figure}

Fig. \ref{Nmedio} shows $\langle N\rangle$ as a function of $G$ for three values of $P$.
It shows that $\langle N\rangle$ linearly increases with $G$, with weaker dependence on $P$.
As $G$ increases by $10$ units, $\langle N\rangle$ typically increases by $1.1$-$1.3$ units.

We also calculated the average number of hops executed by the adatoms after the last detachment
from a terrace border, $\langle N_{last}\rangle$.
If there is no detachment during the adatom diffusion, then the total number of hops is
considered in this average.
The values of $\langle N_{last}\rangle$ are shown in Table \ref{table1} and plotted in Fig. \ref{Nmedio}.
The data for $P>0$ is clearly below the red fitting line for $P=0$, which means that the increase
of $P$ leads to a significant decrease of $\langle N_{last}\rangle$, for fixed $G$.
This contrasts with the increase of $\langle N\rangle$ with $P$ and means that detachment events are
frequent for $P>0$.

\subsection{Interpretation of the results}
\label{interpretation}

When an adatom detaches from a lateral neighbor, it frequently moves to a terrace, at the
same level or at an upper level; see Fig. \ref{hops}.
After the detachment, there is a significant difference in the adatom
mobility in the LM and in the CV model.
In the CV model, this adatom has the same diffusion coefficient of all other terrace
adatoms with the same neighborhood, independently of the time in which they were deposited.
However, in the LM model, the detached adatom will execute less than $G$ hops;
consequently, it has a smaller mobility than a freshly deposited atom.
In simple words, the detached adatom in LM is a "tired adatom".

In many cases, after the detachment from a terrace border, the adatom moves to the site
where it came from before the attachment, i. e. it begins moving from the same position
where it was before attaching to that border.
In the LM model, this adatom can execute a smaller number of hops in comparison with the
first visit to that position; this is consistent with the decrease of $\langle N_{last}\rangle$
with $P$, as shown in Fig. \ref{Nmedio}.
This helps to explain why $\left( G,P>0\right)$ may produce the same
morphology which is produced by $\left( G',P=0\right)$ with $G'<G$, i. e. with a smaller
number of available hops and irreversible lateral aggregation.

In Fig. \ref{Nmedio}, it is also noticeable that the values of $\langle N_{last}\rangle$ are not very
different for the model parameters that produce the same roughness.
This indicates that the diffusion time after the last detachment, $\langle N_{last}\rangle$,
is correlated with the surface roughness, which tends to decrease
as $\langle N_{last}\rangle$ increases.
Instead, the total number of hops of an adatom, $\langle N\rangle$, is not correlated
with the roughness.
This corroborates the interpretation that, for constant $G$, the increase of the number
of detachment events leads to an increase in the number of "tired adatoms" at the surface,
which increases the surface disorder.

The parameter $\epsilon$ in the CV model has the opposite effect on the surface roughness.
The adatom is not "tired" after the detachment from a terrace step; it has the same dynamics as a
freshly deposited atom.
The role of $\epsilon$ is to hold the adatom in an energetically favorable position for a certain
time (typically $\sim \epsilon^{-\left( n-1\right)}$), where it may be buried by the next deposited layer
(permanent attachment).
As $\epsilon$ increases, positions with small $n$ become less favorable, so that permanent attachment
tends to occur at positions with larger $n$.
This is clearly observed in simulations of submonolayer growth \citep{submonorev}.
The preferential attachment at positions with larger $n$ produces more compact structures and explains the
decrease of roughness with $\epsilon$, for fixed $R$.

The nontrivial results for the LM model are related to the non-Markovian rule for adatom diffusion.
This feature leads to very different diffusion properties when compared to Markovian models.
An example is the non-Markovian random walk studied in Ref. \protect\cite{dasilva2014}
(a generalization of the elephant random walk \citep{schutz2004}), in which a hop to a NN site has a
probability that decreases in time as a power law with exponent $-\lambda$.
In the case of unbiased walks, subdiffusion is observed for $0<\lambda\leq 1$, but for larger
$\lambda$ the walker is stationary, i. e. its characteristic displacement has a finite asymptotic value.
This constrasts with the normal diffusion in the case where a hop has a finite probability to be executed.
In our model of adatom diffusion, the condition of a maximal number of $G$ hops corresponds to an infinitely
rapid time decay of the hopping probability at $t=G$ (i. e. a decay much faster than any continuous function).
Moreover, using a Langevin equation approach, it was shown in Ref. \protect\cite{lapas2015} that a 
system with exponentially decreasing memory kernel in contact with a reservoir may have anomalous
thermalization \citep{lapas2015}; note that this is a case with memory of much shorter range than models
of the type of the elephant random walk \citep{schutz2004}.

On the other hand, the Markovian property must not be viewed as a necessary condition for a model to describe
adatom diffusion on surfaces.
For instance, Ref. \protect\cite{pintus} showed that a coarse-grained non-Markovian model was necessary
to reproduce the motion of adsorbed molecules in microporous materials predicted by molecular dynamics
simulations.
Here, the non-Markovian LM model was actually able to provide results consistent with the CV model,
although the apparently corresponding parameters $\epsilon$ and $P$ have different effects
on the surface roughness.

\section{Conclusion}
\label{conclusion}

We introduced a limited mobility model with long surface diffusion lengths of the adsorbed
atoms and with reversible aggregation to lateral neighbors during its diffusion (LM).
These are the same conditions of the simplest model that describes the simultaneous effects
of deposition and collective surface diffusion in thin film growth (CV).
We investigated the scaling properties of the LM model, compared them with the CV model,
and discussed the effects of the probabilities of detachment from terrace steps.

The maximal number of hops in the LM model, $G$, plays a similar role as the
diffusion-to-deposition ratio $R$ in the CV model, since both parameters are related to
the average diffusion time in terraces.
Indeed, the roughness has a significant decrease as $G$ or $R$ increases.
In short lengthscales, increasing $P$ or $\epsilon$ has the effect of forming longer
straight regions in the terrace boundaries.
For given $R$ and $\epsilon$ in the CV model, it is possible to find pairs $\left( G,P\right)$
which lead to the same evolution of the surface roughness and reasonable agreement between
the corresponding autocorrelation functions.
This shows that a properly defined model with limited mobility can reproduce the surface
fluctuations and correlations a more realistic growth model.
For this reason, it may be interesting to extend limited mobility models to consider mechanisms
such as energy barriers at descending steps or anisotropic diffusion.

However, there is a striking difference between the effects of the detachment probabilities
$P$ and $\epsilon$ on the surface roughness, since the increase of $P$ leads to the increase
of the roughness.
While $\epsilon$ can be related to an activation energy and the temperature
[Eq. (\ref{defD})], a similar interpretation for $P$ becomes difficult.
The nontrivial effect of $P$ is related to the reduced diffusion time of an adatom in the
LM model after it detaches from a step edge, i. e. the number of available hops for this adatom
is smaller than the number of hops available for a freshly deposited adatom.
On the other hand, in the CV model, all atoms with the same neighborhood have the same
diffusion coefficient, independently of their history at the film surface.
Quantitatively, we observe that the surface roughness in the LM model is correlated with the 
average number of hops of an adatom after the last detachment from lateral neighbors,
$\langle N_{last}\rangle$.
Instead, the average number of hops since the adsorption, $\langle N\rangle$,
cannot explain the variations of the roughness.

These results suggest that a limited mobility model with a maximal number of allowed hops will
always provide a limited description of a real growth process dominated by surface diffusion.
One possibility for future work is to improve the design of the limited mobility models with
paramters that may be interpreted in a similar way as those of the collective diffusion models.
For instance, the present model may be extended by allowing the adatom at a terrace to
execute $G$ hops even after detachment from a terrace step; this adatom still holds a memory
on the number of previously executed hops, but the depletion of $\langle N_{last}\rangle$ as
$P$ increases may be suppressed.

\section*{Acknowledgment}

The authors acknowledge support from CNPq, CAPES, and FAPERJ (Brazilian agencies).

\bibliographystyle{unsrt}
\bibliography{CVLM}

\end{document}